\magnification=\magstep2
\def\para{\par\noindent}
\def\sqr#1#2{{\vcenter{\vbox{\hrule height.#2pt
        \hbox{\vrule width.#2pt height#1pt \kern#1pt
          \vrule width.#2pt}
        \hrule height.#2pt}}}}

\newcount\notenumber

\def\note{\advance\notenumber by 1
\footnote{$^{\the\notenumber}$}}
\baselineskip 20pt
\centerline{{\bf Chiral Ordering in the Four-Dimensional XY Spin Glass}}\para
\vskip 0.25cm
 \para S.~Jain,
 \para School of Mathematics and Computing,\para
 University of Derby,\para
 Kedleston Road,\para
 Derby DE22 1GB,\para
 U.K.\para
\vskip 0.25cm
\para E-mail: S.Jain@derby.ac.uk
\para 
\para 
\para 
\para 
\vskip 4.0cm
\para Classification Numbers: 0570J, 6460C, 7510H, 7510N
\para 
\para
\vfill\eject
\para {\bf ABSTRACT}
\para The chiral glass behaviour of the nearest-neighbour random-bond
XY spin glass in
 four dimensions is studied by
Monte Carlo simulations. A chiral glass transition at $T_{cg}=0.90\pm 0.05$
is found by a  
finite-size scaling analysis of the results. The associated chiral
 correlation-length exponent is estimated to be $\nu_{cg}=0.6\pm 0.1$ 
and $\eta_{cg}\sim 0.25$.
The values for the chiral critical temperature and the exponents 
are very similar to those reported recently for
the spin glass transition in this model.
 The results strongly suggest a simultaneous ordering of spin and chirality
in four dimensions.
\vfill\eject
\para 1. {\bf Introduction} 
\para Although it has been known since Villain [1] that frustrated vector spin 
systems possess both reflectional and rotational symmetries, it's only fairly
recently that the chiral glass behaviour of vector spin glasses has been
studied [2-9]. Whereas the continuous rotational symmetry is associated with
the spins, it is the discrete twofold Ising-like reflectional symmetry which
is associated with chirality. Much of the recent work has been motivated by
the suggestion that the chiral glass transition in vector spin glasses belongs
to the same universality class as the Ising spin glass transition. This
would imply that the spin glass transition observed in experiments may, in fact,
be \lq chirality driven\rq\ [7]. Consequently, one would have a chiral glass
phase with broken reflectional symmetry but preserved rotational symmetry.
Recent numerical work by Kawamura [4,7,8] in three dimensions 
would appear to be consistent with
this picture.
\para There is now convincing evidence that both XY [3-6,10] and Heisenberg 
[6-8,11] spin
glasses exhibit conventional spin glass ordering only at zero temperature for
$d=2$ and $3$. Domain-wall renormalization-group studies suggest chiral
ordering for both XY [5] and Heisenberg [7] 
spin glasses in two dimensions also at zero
temperature only. However, the values of the chiral and spin glass 
correlation-length exponents have been found to differ in both cases.  
  Further evidence for a decoupling of the chiral and
phase variables on long length scales has come from various Monte Carlo
studies in 2d [2,3], including very recent work using a vortex representation
 [9].
\para In 3d the results point to a finite-temperature chiral glass transition 
both for XY [4-6] and Heisenberg [6-8] spin glasses (although the case is far from 
convincing for the latter). So it would appear that chiralities and spins
have markedly different behaviour in both $d=2$ and $3$.
\para Very recent Monte Carlo simulations in four dimensions suggest a finite-
temperature spin glass transition for vector spin glasses [12,13]. This would
imply that the lower critical dimensionality for vector spin glasses is less
than four. It is clearly of interest to see whether the difference in the
behaviour of chiralities and spins extends into higher dimensions.
\par In this paper we present the results from Monte Carlo simulations of the
chiral behaviour of the four dimensional random-bond XY spin glass. Using 
finite-size scaling and our earlier results for the spin glass transition [12],
we shall find evidence for a simultaneous ordering of spin and chirality in
four dimensions.
\para In section 2 we define the model and review the finite-size scaling technique
used to analyse the data. Section 3 gives details about the simulations and the
results are presented and discussed in section 4. The conclusion is given in
section 5.
\para 2. {\bf The model}
\para The Hamiltonian for the model is given by 
$$ {\it H} =-\sum_{<i,j>} J_{ij} \cos(
\theta_i - \theta_j),\eqno(1)$$
where $0\le\theta_i\le 2\pi$ for all planar spins $i$ and the summation runs 
over all nearest-neighbour pairs on
  a four dimensional hypercubic lattice with $L\times L\times L\times L$
spins $(L=2,4$ and
 $6)$ with full periodic boundary conditions.
 The 
interactions, $J_{ij}$, are independent random variables selected from a
binary $\pm 1$ distribution. Throughout, the temperature is given in
 units of the nearest-neighbour interaction. We recently reported the results
concerning the spin glass transition for this model [12]. In this publication
we concentrate on the chiral glass transition.

The local chirality, $\kappa_\alpha$, at a plaquette $\alpha$ consisting of
four spins is defined by the scalar [2]
$$\kappa_\alpha=2^{-3/2}\sum'_{\alpha}\hbox{Sgn}(J_{ij})\sin (\theta_i-\theta_j),
\eqno(2)$$
where the summation is performed over a directed (clockwise) closed contour
along the four sides of the plaquette. Clearly, whereas 
$\kappa_{\alpha}=0$ for an isolated unfrustrated plaquette, it's restricted to 
the values $\pm 1$
for frustrated ones. Hence, chirality act as a \lq continuous\rq\ Ising-like
quantity. However, as it's magnitude fluctuates with the
temperature to a certain extent, we work with a root-mean-square value
 given by [2]
$${\overline\kappa} = \sqrt{{1\over N_d}\sum_{\alpha}
[<\kappa_{\alpha}^2>_T]_J},\eqno(3)$$
where $<...>_T$ denotes a thermal average, $[...]_J$ indicates an average
over the disorder and $N_d$, the total number of plaquettes 
for our four dimensional
lattice , is equal to $ 6 L^4$.
\para At a chiral glass transition one expects the chiral glass susceptibility,
$\chi_{cg}$, to diverge, where
$$\eqalignno{\chi_{cg}& = {1\over N_d}\sum_{\alpha,\beta}[<\kappa_\alpha\kappa_
\beta>^2_T]_J\cr
     &=N_d q^{(2)}_{cg},&(4)\cr}$$
and here the summation is with respect to all plaquettes $\alpha$ and $\beta$.
\para The chiral glass order parameter, $q^{(2)}_{cg}$, defined above can
be written in terms of the overlap between two replicas $1$ and $2$, viz

$$q^{(2)}_{cg}=[<q^2>_T]_J,\eqno(5)$$
where
$$q={1\over N_d}\sum_{\alpha}\kappa_{\alpha}^1\kappa_{\alpha}^2.\eqno(6)$$
In order to improve the analogy with Ising spins, we follow Kawamura [2] and
work in the simulations with the reduced chiral glass susceptibility
$${\widetilde{\chi}_{cg}}=\chi_{cg}/\overline\kappa\ ^4.\eqno(7)$$
Another key quantity studied in the simulations is the
 dimensionless Binder parameter
 defined by [14]
$$g_{cg}(L,T)={1\over 2}\bigg[3-{{q_{cg}^{(4)}}\over {(q_{cg}^{(2)})^2}}\bigg]\eqno(8)$$
and here
$$q_{cg}^{(4)}=[<q^4>_T]_J.\eqno(9)$$

The Binder parameter is expected to scale as [14]
$$ g_{cg}(L,T)={\overline g_{cg}}(L^{1/\nu_{cg}}(T-T_{cg})),\eqno(10)$$
where $T_{cg}$ and $\nu_{cg}$ are the chiral glass critical temperature and
correlation-length exponent, respectively. The value of $T_{cg}$ can be located
 by using the fact that the scaling function, $\overline g_{cg}$, 
and, hence, also $g_{cg}$ are
independent of $L$ at the transition temperature. 
The chiral glass correlation-length
exponent then follows from a one-parameter scaling fit of the data for $g_{cg}$.
\para The finite-size scaling form for the reduced chiral glass susceptibility 
is given by 
$${\widetilde{\chi}_{cg}(L,T)}=L^{2-\eta_{cg}}
{\overline{\widetilde{\chi}}}_{cg}
(L^{1/\nu_{cg}}(T-T_{cg})).\eqno(11)$$ 
Here $\eta_{cg}$ is the chiral critical-point decay exponent 
and $\overline{\widetilde{\chi}}_{cg}$
the scaling function.
\par The finite-size scaling form for the analogous spin glass quantities can
be found in [12].
\para 3. {\bf Simulations}
\para We now discuss the computer simulations (Jain [12] should
be consulted for further technical details).
 In order to ensure that
equilibrium has been achieved in the simulations, we use the technique of
Bhatt and Young [14] whereby chiral glass correlations $(q_{cg}^{(2)}$ and $q_{cg}^{(4)})$
computed from two replicas at the same time are required to
 agree with those from one 
replica at two different times.
\para At low temperatures the chiral degrees of freedom
 were more difficult to equilibrate than their spin counterparts, requiring
 approximately twice as many Monte Carlo steps. As a result,
the lowest temperatures studied were $T=0.8 (L=6), 0.7 (L=4)$ and $0.3 (L=2)$.
We took disorder averages over many pairs of independent samples: $100\sim 120 (L=6),
100\sim 250 (L=4)$ and $250\sim 700 (L=2)$ for each temperature. Most of
the computational time (about 200 hours of CPU time on a Cray J932) was
 taken up by the simulations for $L=6$.
 (The processors of the Cray J932 are roughly only half the speed of 
those of the Cray YMP.) It's estimated that an additional 1000 hours of CPU time
would be required to obtain reliable data on a larger lattice such as $L=8$. All
computational studies performed to-date, including the present one, on four-dimensional
spin glasses have been restricted to small lattices with $L\le 6$ [12-14].
\para 4. {\bf Results}
\para In this section we present the results. 
\para Figure 1 shows a plot of the reduced chiral glass susceptibility against
temperature for the three different lattices considered in this work. The 
statistical error-bars have been estimated from the sample-to-sample fluctuations
and, in most cases, are smaller than the size of the data points.
\par We see that the value of $\widetilde{\chi}_{cg}$ remains quite low until
$T\approx 1.0$ and then increases rapidly with the system size.
In fact, whereas for low temperatures ${\widetilde\chi}_{cg}$ is an
increasing function of $L$, for higher temperatures it appears to
actually decrease with increasing $L$. This suggests that the scaling regime
is probably quite narrow. Very similar behaviour was found recently
by Kawamura in 3d [4].

\para In figure 2(a) we show
a plot of $g_{cg}$ against the temperature for $0.4\le T\le 1.5$. 
The behaviour of the Binder parameter for high temperatures $T\ge 1.0$ 
indicates a disordered chiral phase. For $T\le 1.0$ there is a sharp increase
in $g_{cg}$,
indicating a build up of chiral correlations. The increase is more noticeable for
the larger lattices. We notice that for $T\ge 1.0\ g_{cg}$ assumes neagtive
values. A negative Binder parameter has also been seen for both XY [4]
and Heisenberg [8] spin glasses in 3d. This feature would appear to be a 
consequence of the fact that $\kappa_{\alpha}=0$ on unfrustrated plaquettes
even in the {\it ordered} state.
\par Figure 2(b) displays on a much expanded scale
 the data in figure 2(a) in the vicinity of
the important temperature region $(0.8\le T\le 1.0)$. We see that the 
curves appear to
 intersect
at $T_{cg}\approx 0.9$. Furthermore, below $T_{cg}$ the values corresponding to 
$L=6$ are consistently
above those for $L=4$ and the curves clearly splay out.
 We estimate the chiral glass transition temperature to be
$T_{cg}=0.90\pm 0.05$. This value is very close to the spin glass transition
temperature $(T_{sg}=0.95\pm 0.15)$ we reported recently for this model [12].
Our conclusion is based on the results for small lattices and it is highly
desirable to obtain additional data for larger lattices to confirm our findings.
\para Further evidence for a finite-temperature transition comes from a
scaling plot of the Binder parameter. By varying
the value of $\nu_{cg}$ and considering values $0.85\le T_{cg}\le 0.95$,
we estimate the chiral correlation-length exponent to be $\nu_{cg}=0.6\pm 0.1$.
It should be noted that the error-bar quoted here is simply an estimate that
demarcates the range of values beyond which the data do not scale well.
Figure 3 shows the data for $g_{cg}$
 against $L^{1/\nu_{cg}}(T-T_{cg})$ with $T_{cg}=0.90$ and
$\nu_{cg}=0.6$. This scaling plot is far better than the corresponding
plot for $g_{sg}$ (see figure 4 in [12]) and the data (including those for
$L=2$) would appear
to scale particularly well near $T_{cg}$. Once again, the value of $\nu_{cg}$
would appear to be very similar to that of $\nu_{sg}=0.70\pm 0.10$ [12].   
\para We have appreciable uncertainty in the values of
both $T_{cg}$ and $\nu_{cg}$. Furthermore, the increase in the reduced chiral
glass susceptibility for $T\le 1.0$ is extremely sharp. As a consequence, 
the critical region is very narrow. To obtain an estimate for the decay
exponent, $\eta_{cg}$, we tried various different possible values of $T_{cg}$
and $\nu_{cg}$. The best such scaling plot is shown in figure 4 where we
display $\widetilde{\chi}_{cg}/L^{2-\eta_{cg}}$ against $L^{1/\nu_{cg}}(T-T_{cg})$
with $T_{cg}=0.85, \nu_{cg}=0.8$ and $\eta_{cg}=0.25$.

Clearly, the $\widetilde{\chi}_{cg}$ data for $L=2$ do not scale
 all that well, especially for the higher temperatures. Nevertheless, we note
that our estimate for $\eta_{cg}$ is not incompatible with the value
 for $\eta_{sg}$ 
found earlier [12].
\vfill\eject
\para 5. {\bf Conclusion}
\para To conclude, we have presented the results of a Monte Carlo simulation
of the chiral glass behaviour of the four dimensional random-bond XY spin glass
on small lattices $(L\le 6)$.
By means of a finite-size scaling analysis of the data, we have estimated
both the chiral transition temperature and the critical exponents. The chiral
glass values are very
similar to their spin glass counterparts
 for this model ($T_{cg}\approx T_{sg},
\nu_{cg}\approx\nu_{sg}$ and $\eta_{cg}\approx \eta_{sg})$. Hence, our results strongly suggest a simultaneous
ordering of spin and chirality in four dimensions. 
\par Further data on larger lattices
($L>6$) is required to confirm both the transition temperature and the critical exponents.

\para {\bf Acknowledgement}
\para The simulations were performed on a Cray J932
 at the Rutherford Appleton 
Laboratory through an Engineering and Physical Sciences Research Council 
(EPSRC) research grant (Ref: GR/K/00813).  
\vfill\eject 
\para FIGURE CAPTIONS
\vskip 1cm
\para Figure 1

\para A plot of the reduced chiral glass susceptibility, $\widetilde{\chi}_{cg}$, 
against the
temperature for $L=2, 4$ and $6$ (see equations $(4)$
 and $(7)$ in the text). The lines are just to guide the eye.
\vskip 1cm

\para Figure 2(a)

\para A plot of the Binder parameter $g_{cg}$ against the
 temperature for $L=2, 4$ and $6$. 
\vskip 1cm

\para Figure 2(b)

\para Same as figure 2(a) but on a much expanded scale around the interesting
temperature region $(T\approx 0.9)$. The lines are just to guide the eye.

\vskip 1cm

\para Figure 3

\para A scaling plot of the Binder parameter $g_{cg}$ versus
 $L^{1/0.6}(T-0.9)$. The line is just a guide to the eye.
\vfill\eject
\para Figure 4

\para A scaling plot of $\widetilde{\chi}_{cg}/L^{2-\eta_{cg}}$ against
$L^{1/\nu_{cg}}(T-T_{cg})$ with $T_{cg}=0.85, \nu_{cg}=0.8$ and $\eta_{cg}=0.25$
(see equation (11) and the text). The line is just a guide to the eye.
\vfill\eject
\para REFERENCES
\item {[1]} Villain J 1977 J. Phys. C: Solid State Phys. {\bf 10} 4793;
\item {} 1978 J. Phys. C: Solid State Phys. {\bf 11} 745
\item {[2]} Kawamura H and Tanemura M 1985 J. Phys. Soc. Jpn. {\bf 54} 4479; 1986
J. Phys. Soc. Jpn. {\bf 55} 1802; 1987 Phys. Rev. B{\bf 36} 7177
\item {[3]} Ray P and Moore M A 1992 Phys. Rev. B{\bf 45} 5361
\item {[4]} Kawamura H 1992 J. Phys. Soc. Jpn. {\bf 61} 3062; 1995 Phys. Rev. B{\bf 51}
12398 
\item {[5]} Kawamura H and Tanemura M 1991 J. Phys. Soc. Jpn. {\bf 60} 608
\item {[6]} Kawamura H 1995 Computational Physics as a 
New Frontier in Condensed Matter Research 209
\item {[7]} Kawamura H 1992 Phys. Rev. Lett. {\bf 68} 3785
\item {[8]} Kawamura H 1995 J. Phys. Soc. Jpn. {\bf 64} 26
\item {[9]} Bokil H S and Young A P 1996 J. Phys. A: Math. and Gen. {\bf 29} L89
\item {[10]} Banavar J R and Cieplak M 1982 Phys. Rev. Lett. {\bf 48}
832
\item {} McMillan W L 1985 Phys. Rev. B {\bf 31} 342
\item {} Jain S and Young A P 1986 J. Phys. C: Solid State Phys. {\bf 19} 3913
\item {[11]} Olive J A, Young A P and Sherrington D 1986 Phys. Rev. B {\bf 34}
 6341
\item {[12]} Jain S 1996 J. Phys. A: Math. and Gen. {\bf 29} L385
\item {[13]} Coluzzi B 1995 J. Phys. A: Math. and Gen. {\bf 28} 747
\item {[14]} Bhatt R N and Young A P 1988 Phys. Rev. B {\bf 37} 5606
\end